# Stability of ferroelectric bubble domains

Vivasha Govinden[1*], Suyash Rijal[2*], Qi Zhang[1†], Yousra Nahas[2], Laurent Bellaiche[2], Nagarajan Valanoor[1], Sergei Prokhorenko[2†]

[1] *School of Materials Science and Engineering, University of New South Wales, Sydney, NSW 2052, Australia*

[2] *Physics Department and Institute for Nanoscience and Engineering, University of Arkansas, Fayetteville AR 72701, USA*

[*] these authors contributed equally

† corresponding authors:
Qi Zhang: peggy.zhang@unsw.edu.au

Sergei Prokhorenko: sprokhor@uark.edu

## Abstract

Nanoscale ferroelectric topologies such as vortices, anti-vortices, bubble patterns etc. are stabilized in thin films by a delicate balance of both mechanical and electrical boundary conditions. A systematic understanding of the phase stability of bubble domains, particularly when the above factors act simultaneously, remains elusive.  Here we present first-principle-based simulations in combination with scanning probe microscopy of ultrathin epitaxial (001) PbZr$_{0.4}$Ti$_{0.6}$O$_3$ heterostructures to address this gap. The simulations predict that as-grown labyrinthine domains will transform to bubbles under combinations of reduced film thickness, increased mechanical pressure and/or improved electrical screening. These topological transitions are explained by a common fundamental mechanism. Namely, we argue that, independently of the nature of the driving force, the evolution of the domain morphology allows the system to conserve its original residual depolarization field. Thereby, the latter remains pinned to a value determined by an external or built-in electric bias. To verify our predictions, we then exploit tomographic atomic force microscopy to achieve the concurrent effect of reducing film thickness and increased mechanical stimulus. The results provide a systematic understanding of phase stability and demonstrate controlled manipulation of nanoscale ferroelectric bubble domains.

Topological structures in ferroelectric systems have recently drawn immense attention. For example, topological configurations such as flux closures [1-4], vortex pairs [5-8], skyrmions [9, 10] or nanoscale bubble domains [11, 12] have led to emergent properties including high conductivity [13], chirality [14, 15], negative capacitance [16, 19], and giant electromechanical response [11]. These attributes make polar topologies promising candidates for various device applications, especially where they can be deterministically controlled.

The stability of dipolar topological patterns is subject to a subtle balance of mechanical and electrical boundary conditions [11-20]. For instance, in the case of polar vortices we now know there is a fine balance between long-range electrostatic and elastic versus short-range polarization gradient interactions that determine vortex stability. Damodaran et al. [21] exploited this balance to practically realize phase co-existence where interconversion from the topological vortex phase to a ferroelectric a1/a2 phase resulted in giant piezoelectric and non-linear optical responses [22]. More than a decade ago, Balke et al. reported the controlled creation of a ferrotorroidal order parameter by exploiting the trailing fields of a scanning probe tip [23]; this approach was most recently used to create a vortex-anti vortex pair by Kim et al. in an otherwise topologically trivial ferroelectric canvas [5].

In contrast, understanding of the phase stability of polar bubbles in ferroelectric films still has open questions. Polar bubbles in $PbZr_{0.2}Ti_{0.8}O_3/SrTiO_3/PbZr_{0.2}Ti_{0.8}O_3$ heterostructures are homotopy equivalent to the polar skyrmions reported in $PbTiO_3/SrTiO_3$ (PTO/STO) superlattices [24] but can be achiral, in contrast to the PTO-based ones [25]. Consequently, these two types of spherical topological defects appear under different external conditions [9, 26]. Experimentally, stabilizing bubble domains is not a trivial task; even a modest change in the external boundary conditions triggers a rapid phase transition to labyrinthine or monodomains [11, 26]. The bubble domains have significant importance for technology as (*i*) they are a precursor to the much sought-after electrical skyrmion [24], (*ii*) promise giant piezoelectric response [11], (*iii*) can be realized in free-standing films [27] and (*iv*) possess dynamic intrinsic conductivity [28] that could be harnessed in neuromorphic computing [29]. Although we have captured the individual effects of factors such as thermal and bias profiles [29], mechanical pressure [26], thickness [30] and screening [31], exploiting these properties for a topological device requires knowledge of what happens when several of these external tuning parameters act in tandem.

This report provides an in-depth understanding of the phase stability of bubble domains in ferroelectric thin films under such simultaneous stimuli. Starting with first-principle-based effective Hamiltonian simulations as guide, we investigate the simultaneous effects of film thickness, external mechanical force and screening conditions on the topological defect transition mechanisms in epitaxial (001) PbZr$_{0.4}$Ti$_{0.6}$O$_3$ (PZT) thin films. Our simulations show that the mechanical stress normal to the film's surface, the decreased films thickness as well as enhanced electric screening have the same effect: the thinning and disconnecting of the domain pattern. We further show that, in all cases, the resulting sequence of labyrinth-bubble-monodomain transitions [29] can be explained by an unusual "conservation law" – an intrinsic tendency of the ferroelectric film to preserve its original depolarizing field. These predictions are then verified experimentally by exploiting the newly developed tomographic atomic force microscopy (TAFM) technique [32, 33] on PbZr$_{0.4}$Ti$_{0.6}$O$_3$/La$_{0.67}$Sr$_{0.33}$MnO$_3$//SrTiO$_3$ (PZT-single layer) and PZT/STO/PZT/LSMO//STO (PZT-sandwich) films respectively under a variety of mechanical forces and at different thicknesses. Under compressive stress (applied through the TAFM tip), as-grown labyrinthine-shaped domains transform to disconnected labyrinths (PZT-single layer) or nanoscale bubbles (PZT-sandwich) dependent on the screening conditions. On the other hand, TAFM milling, i.e., which reduces thickness, results in bubble domains in both film configurations. The experiments confirm that the stability of these different ferroelectric topologies is ultimately linked to how mechanical stress, screening or thickness of the system modulate the residual depolarizing field [34, 35] in the thin film heterostructure.

We begin with the effective Hamiltonian simulations of PbZr$_{0.4}$Ti$_{0.6}$O$_3$ ultrathin films [20] that allow us to understand how each of the boundary conditions govern the topological defect structures. For all cases we assume the films are (001) oriented with an imposed in-plane biaxial epitaxial strain of -2% [36] and a built-in bias field $E_b$ of 0.18 V/nm. Full details of simulations are provided in the Materials and Methods section of the Supplemental Material. Figure 1a-e shows various simulated domain patterns for the investigated range of external boundary conditions.

In the virgin state, our simulations predict that the domains have a disconnected labyrinthine-like morphology (Fig. 1a). Upon applying an increasing homogenous stress ($\sigma$) perpendicular to the film surface, the labyrinthine domains progressively disconnect, transform into bubbles and eventually disappear to yield a monodomain state. Fig. 1b demonstrates the simulated domain pattern precursor to the bubble phase at $\sigma$ = 1 GPa compressive stress. The same sequence of labyrinth-bubble-monodomain transitions is predicted by simulations when either enhancing the screening $\beta$ (e.g., from $\beta$=0.86 in Fig. 1a to $\beta$=0.88 Fig. 1c) or decreasing the film thickness h (e.g., from $h = 8$ unit cells (u.c.) in Fig. 1b to $h$=7 u.c. in Fig. 1d). Moreover, this chain of domain pattern transformations is not altered when tuning several external parameters. All factors work hand-in-hand and adjusting one variable allows to ease the transitions triggered by the other. For example, ultra-fine bubbles (Fig. 1e) are seen with just half the applied stress ($\sigma$=0.5 GPa) when the PZT thickness is reduced ($h = 7$ u.c) and screening conditions are enhanced ($\beta$=0.88).

To understand the physics of the described transitions, we first recall that the non-trivial domain topologies in ultra-thin PZT films fall within the family of universal self-organized patterns commonly found in phase separation kinetics [29]. The link is granted by the condition of the bias field ($\boldsymbol{E_b}$) compensation by the residual depolarizing field ($\boldsymbol{E_d}$) that holds up to the monodomain state (i.e., it is valid for the labyrinthine and bubble patterns) (see Supplementary Material and Fig. S1). Here, since the bias magnitude $E_b$ always remains fixed, the $E_d = E_b$ relation simply means that the residual depolarizing field is conserved under any change of external parameters.

The signatures of conserved depolarizing field are seen in all of our present simulations. For example, we find that, under mechanical loading, the out-of-plane polarization $P_z$ only weakly responds to the applied stress despite a strong suppression of the average magnitude of local electric dipoles $|\boldsymbol{d}|$ (Fig. 1f). This assures a steady level of the bound charge at the interfaces thereby guaranteeing a constant $E_d$ magnitude. Microscopically, the only way to maintain $P_z$ under reducing $|\boldsymbol{d}|$ is by switching more dipoles along the bias. Such switching progressively disconnects labyrinth domains and leads to bubble formation when σ increases. The scenario of $\beta$-engineering is somewhat opposite. Unlike stress, the screening strength directly affects the depolarizing field and to compensate for a linear drop of $E_d$ with increasing $\beta$ (Supplementary Information Fig. S1) the polarization grows as $P_z \sim 1/(\beta_{SC} - \beta)$ (Fig. 1g). Here, $\beta_{SC}$ denotes the screening strength at which the residual depolarizing field vanishes. In the linear approximation, $\beta_{SC}$ also corresponds to critical screening at which the monodomain pattern forms. At the same time, our simulations show that $|\boldsymbol{d}|$ remains basically invariant up to the bubble-monodomain transition (Figure 1g). Therefore, the growth of $P_z$ is, once again, achieved by switching the dipoles so as to increase the area of the domain polarized along the bias. Finally, for the thickness driven transitions both microscopic mechanisms of depolarizing field conservation come into play. Namely, when the thickness is reduced, the dipole magnitude $|\boldsymbol{d}|$ decays (Fig. 1h) just as in the case of increasing mechanical force. Such weakening of $|\boldsymbol{d}|$ is due to an enhanced role of fluctuations at reduced thickness and dimensional confinement. Moreover, increasing fluctuations enhances the dielectric permittivity ε and, in order to conserve $E_d = P_z/\varepsilon$, the polarization is forced to increase when the thickness is reduced (Fig. 1h). The latter is similar to the enhanced screening scenario.

Testing the validity of the simulations requires an experimental method that facilitates the simultaneous application of electric field and mechanical pressure, change in thickness as well as screening conditions. Fortunately, recent advances in scanning probe microscopy (SPM) techniques [32, 37-39] allow us to image, characterize, as well as manipulate the topological domain patterns under a variety of external boundary conditions with a resolution of up to a few nanometers [6, 26, 37, 40-41]. In this work we exploit the recently developed TAFM [32-33]. Briefly, a stiff scanning probe with a large force constant is employed such that the applied force is sufficient to mechanically remove a layer of the thin film during scanning and hence access a fresh tomographic surface [32] (A tomographic surface is defined as the surface acquired after milling off its above layers). As milling takes place, the thickness of the residual layer decreases (see Supplementary Material Fig. S2). To incorporate the effect of screening, we compared two candidate (001) epitaxial heterostructures- viz. (i) PZT-single layer and, (ii) PZT-sandwich (synthesis details found in Supplemental Material). The TAFM calibration parameters for both systems can also be found in Supplementary Information. Suffice to state here, that a critical force of 7 μN is needed to mill the layers in the PZT-sandwich and, and typically each 7 μN scan causes approximately 4 ± 2 nm depletion. On the other hand, a force of 15 μN is required to mill through the PZT-single layer. Each TAFM scan is followed by an individual piezoresponse force microscopy (PFM) scan to acquire the domain features at the specific surface depth.

We start with the effects of mechanical force and film thickness on the domain patterns of the PZT-single layer. Figure 2a reveals the single layer possesses labyrinth domains in the as-grown state with a width of approximately 50 ±10 nm; care was taken to acquire this image under 100 nN force, which has negligible influence on the domain pattern [26]. The full images of PFM amplitude and phase are given in Supplementary Information Fig. S3. When the scanning force is increased by an order of magnitude or two (i.e., from 100 nN to ⩾1 μN), the as-grown poled-up labyrinthine domains break down into smaller patterns domain transitions, as shown in Fig. 2b. Note that no topography change is observed, suggesting that the transitions are not correlated with either surface damage or ferroelastic switching. Under 15 µN, the wide labyrinth domains shrink into either nanoscale bubble, or disconnected labyrinthine domains (with a length and width of approximately 100 nm and 20±5 nm, respectively). Repeated scans at 15 µN cause PZT milling and when 4 nm of PZT is milled, domains further shrink into a bubble domain only configuration (Fig. 2c).

On the other hand, as-grown domain patterns in PZT-sandwich film are narrower labyrinthine domains with a width of 30±5 nm (Fig. 2d, full images are given in Supplementary Information Fig. S4). Given that both PZT-single layer and PZT-sandwich film have similar out-of-plane lattice parameters (X-ray diffraction and structural data given in Supplementary Material Fig. S5) and comparable screening conditions from electrodes (as they have same LSMO thickness, i.e., 40 nm), the contrast in domain size between two films arises from the differences in their respective electrical boundary conditions (i.e., internal screening) induced by the presence (or absence) of the STO spacer [11]. Under a relatively smaller force (i.e., ~5 µN), the as-grown narrow labyrinthine domain patterns change into nanoscale bubble-like domains. The average planar domain size is approximately 10 nm as shown in their amplitude images (Fig. 2e).

The above data agree with the simulations under mechanical force (Fig. 1a-b), demonstrating a simple route of acquiring stable nanoscale bubble domains with the application of external mechanical force on as-grown labyrinthine domains. Under the external mechanical force applied using the SPM probe, the film is subject to an out-of-plane compressive strain [37]. As discussed in the simulations (Fig. 1f), this compressive strain drives the polarization switching to increase the area of the matrix domain. An increased force to 7 µN results in milling off 4 nm PZT in the PZT-sandwich layer, as well. Although the ultimate topological structure in both systems corresponds to bubble-like domains (Fig. 2c and Fig. 2f), the evolution pathway and driving forces are markedly different. Particularly for the sandwich film, we found the bubbles seen in Fig. 2e undergo a transition to short labyrinths with reducing thickness. This evolution pathway is addressed next.

TAFM of the sandwich film under progressive mechanical loading was carried out to reveal the evolution of the ferroelectric topology when reduced film thickness and mechanical force act simultaneously. Milling precise layer-by-layer of the film surface materials using TAFM conditions followed by PFM imaging facilitates a direct comparison of the domain patterns of the as-grown surface with tomographic surfaces at different depths. The PFM amplitude figures of PZT -sandwich film under 1 µN scanning (Fig. 3a), after 4 µN scanning (Fig. 3b), and after 7 µN scanning for one to four times (Fig. 3c-f), are compared in Fig. 3. The dashed rectangular box locks in the reference location. Three typical regions (I, II and III) are selected to facilitate the understanding of domain evolutions as a function of scanned surface depth (as highlighted in Fig. 3a and schematically illustrated in Fig. 3g).

Below 4 µN mechanical force, no milling takes place and the domain patterns (Fig. 3a and 3b) obtained correspond to those on the film surface. As-grown labyrinthine domains (e.g., in Regions I and II) break down into nanodomains when the force increases from 1 µN to 4 µN. With the scanning force further increased to 7 µN, the film starts to be milled off and the tomographic surface reaches the depth of 4±2 nm (Fig. 3c). At this stage, some nano domains remain as fuzzy bubble configurations with even smaller planar sizes of approximately 3-10 nm (e.g., in Region I), while newly nucleated nanodomains are found in Region II (indicated by white arrows). Thus, when the scanning surface is above the STO spacer, domains evolve from labyrinthine towards smaller nanoscale domains due to the decreasing thickness of the PZT layer. With the removal of STO spacer (i.e., from Fig. 3d), domains evolve significantly due to the variation of the screening conditions [11]. Particularly in Region III (dashed oval in Fig. 3a-f), new labyrinthine-like domains are formed in the bottom PZT layer (Fig. 3d-f) which are completely different from the ones at film surface (i.e., no domains are observed in Region III surface, Fig. 3a).

In addition, we highlight most of the nanodomains from Region II in 4 nm depth layer are absent in the next 8 nm depth layer (Fig. 3d). Also, some nanodomains in surface layer are found to disappear in the 4 nm tomographic surface layer. This suggests such nanodomains domains only exist within the 4 nm thickness. All such nanodomains also exhibit fuzzy wall features consistent with bubble domains [11]. Therefore, using TAFM technique, the domain patterns can be changed along labyrinthine → disconnected labyrinthine ↔ nanoscale bubble domains pathway through applying the external force and changing residual film thickness. The results confirm the universal breakdown of labyrinthine (Fig. 3a) into bubble domains (Fig. 3b) under external mechanical pressure, in alignment with the numerical simulations of Fig. 1a-b. When this mechanical pressure is accompanied by milling, i.e., the effective ferroelectric thickness is being reduced, the bubble domains become finer (Fig. 3c), confirming thickness reduction in tandem with external compressive mechanical force, results in finer bubbles (Fig. 1d). Hence domains with specific topological configurations (or transitions) can be deterministically induced by precise control of the depth penetration of the mechanical force, highlighting the opportunity for tailored external condition within TAFM.

These above observations in conjunction with our previous work now allow us to provide a complete understanding of the phase stability of ferroelectric bubble domains:

- Topology of polar domains is determined by the balance of the depolarizing and bias fields. The former can be indirectly tuned by adjusting the thickness of the PZT layers, changing the screening conditions or exerting an out-of-plane mechanical stress. Modifying any of these parameters shifts the balance which triggers the domain pattern evolution needed to restore the residual depolarization.
- Ferroelectric bubble domains are an intermediary topological state between the stripes/labyrinth domains and the monodomain state [11, 29]. This "Goldilocks" state is most readily realized experimentally by the insertion of an ultrathin dielectric spacer layer between two c-axis oriented epitaxial ferroelectric layers which are under modest in-plane compressive strains.
- Local mechanical pressure can be exploited to trigger topological defect transitions; the sequence of phase stability goes from labyrinth→bubbles→monodomain under a compressive strain applied along the polarization direction. Under the influence of mechanical pressure, as-grown bubbles can be erased to form a monodomain state [26], analogous to mechanical writing of polarization [37]. On the other hand, mechanical pressure can also trigger the breakdown of as-grown labyrinthine domains into bubbles.
- Thermal quenching leads to either configuration depending on the strength of the applied or built-in electric fields. Thermal quenching as a route to capture non-equilibrium states, results into labyrinthine domains (or bubbles) under the absence (or presence) of an electric field [29].
- Interface effects can be further exploited to engineer topological defect transitions. Varying the thickness of the ferroelectric and/or dielectric layer in a ferroelectric/dielectric/ferroelectric heterostructure modifies the strain, screening conditions as well as the bare depolarizing field and thus plays a significant role in the resulting domains. Under varying thicknesses of these layers, a transition from labyrinthine to bubble domains can be triggered [30-31]. A similar domain transition occurs upon milling through the ferroelectric layer.

## Acknowledgments

The research at University of New South Wales (UNSW) was supported by DARPA Grant No. HR0011727183-D18AP00010 (TEE Program), partially supported by the Australian Research Council Centre of Excellence in Future Low-Energy Electronics Technologies (project number CE170100039) and funded by the Australian Government. Q. Z. acknowledges the support of a Women in FLEET Fellowship. The research at University of Arkansas is also supported by the Vannevar Bush Faculty Fellowship (VBFF) Grant No. N00014-20-1-2834 from the Department of Defense and Arkansas Research Alliance.

# List of figures

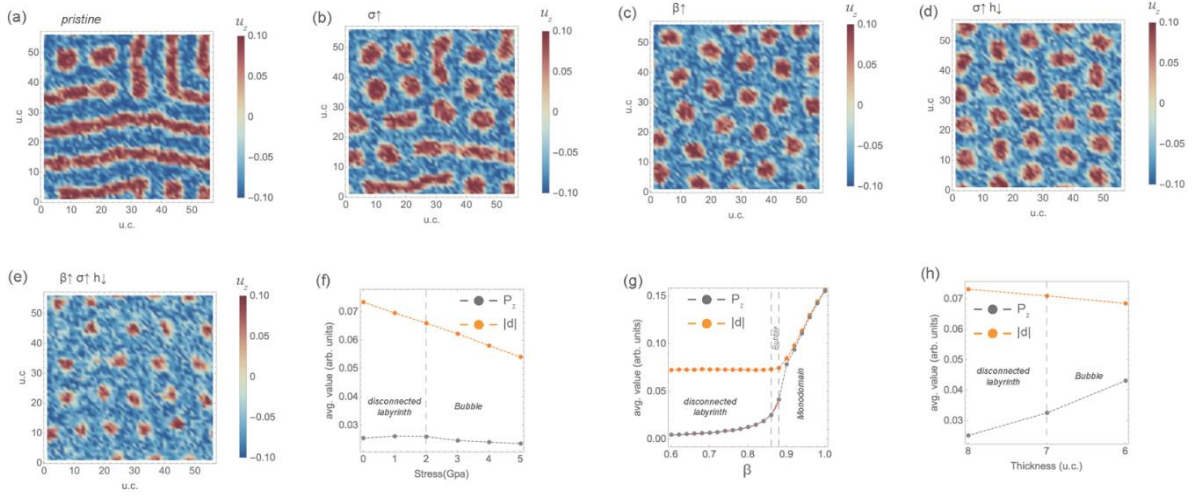

**Figure 1.** Effective Hamiltonian simulations of domain structure for different PZT film thickness $h$, screening $\beta$ and the out-of-plane stress $\sigma$. (a) Labyrinth domains at $\sigma = 0$ GPa, $\beta$=0.86 and for $h$=8 u.c. (b) Disconnected labyrinth pattern with bubble domains at $\sigma = 1$ GPa, $\beta$=0.86, and $h$=8 u.c. Simulated arrays of bubble domains for (c) $\sigma = 0$ GPa, $\beta$=0.88, $h$=8 u.c and (d) $\sigma = 1$ GPa, $\beta$=0.86, $h$=7 u.c. (e) Ultra-fine polar bubbles at σ = 0.5 GPa, β=0.88, h=7 u.c. Dependence of the average dipole magnitude |d| and the out-of-plane polarization $P_z$ on (f) applied stress σ for $h$=8 u.c and $\beta$=0.86 (g) screening $\beta$ for $h$=8 u.c. and $\sigma$ =0 GPa and (h) thickness $h$ for $\sigma$ =0 GPa and $\beta$=0.86. Dashed lines indicate transitions. Solid red line in (g) shows the fit $P_z = C/(\beta_{SC} - \beta)$ with fitting parameters $C$ and $\beta_{SC}$. Panels (a-e) show the

middle-(001)-plane cross-sections of a supercell where red (blue) colors indicate dipoles pointing along [001] ([00$\bar{1}$]) pseudo cubic direction.

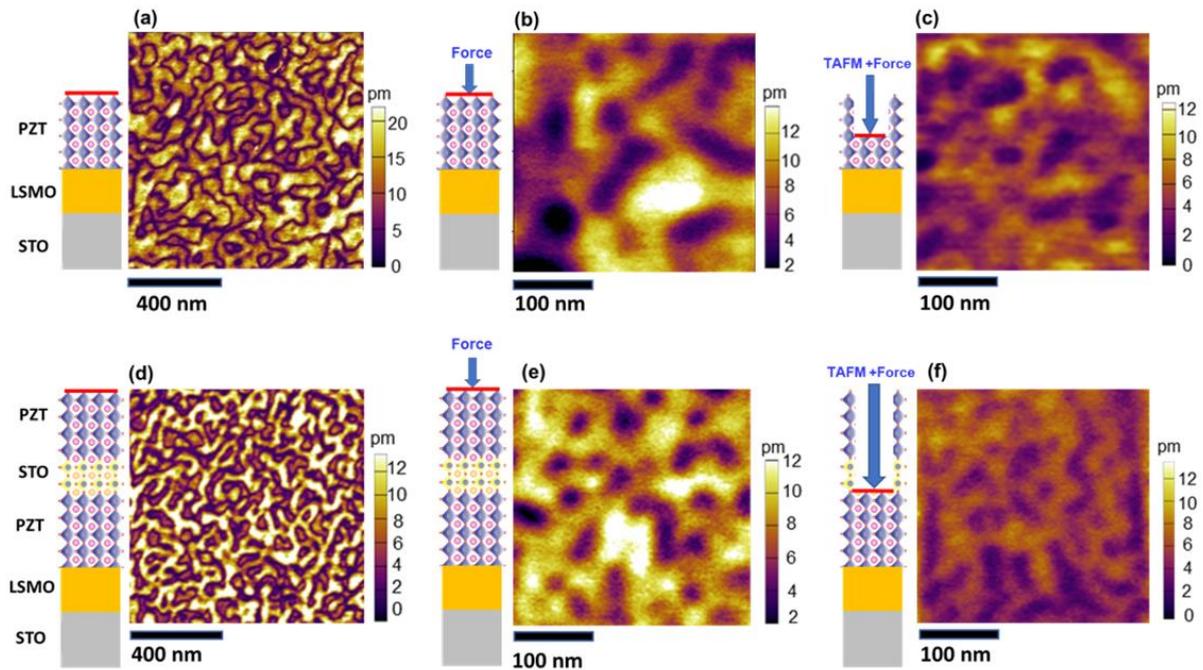

**Figure 2.** (a)-(c) Domain patterns of in PZT-single layer under various conditions: (a) As-grown large labyrinthine domains. (b) Smaller labyrinthine domains are obtained under mechanical force. (c) Bubble domains are formed when ~4 nm of PZT is milled off. (d)-(f) domain patterns of in PZT/STO/PZT-sandwich film under various conditions: (d) As-grown (thinner) labyrinthine domains. (e) Smaller labyrinthine domains and bubbles are obtained under mechanical force. (f) After milling through first PZT layer, film configuration is similar to (b) and domain patterns also look alike (small labyrinthine + bubble domains). After further milling through bottom PZT, bubble domains are formed. The scanning surfaces are indicated by the red line in the corresponding schematics. Note the schematics are a guide and do not reflect the real unit cell numbers.

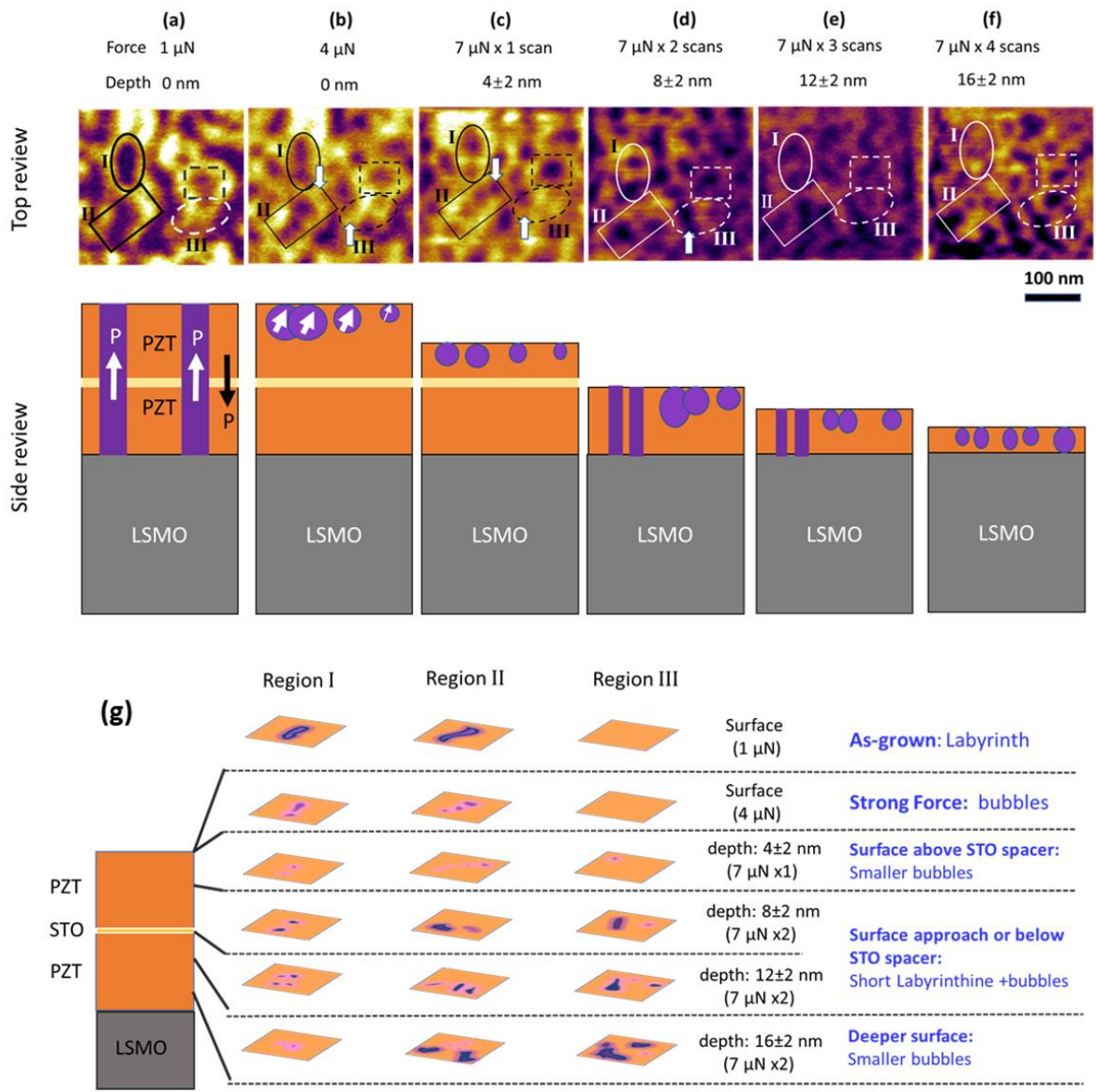

**Figure 3.** Domain configuration as a function of force and depth: (a) 1 µN, (b) 4 µN, (c) 7 µN after 1 scan, (d) 7 µN after 2 scans, (e) 7 µN after 3 scans, and (f) 7 µN after 4 scans. The white box showing the reference domain area. With the force increases, the domains first changes into bubble -like domains, and then labyrinthine domains were observed again at a lower depth. (g) Schematic of domain evolutions as a function of force and depth in Region I, Region II and Region III in (a).

# Supplemental Material: Stability of ferroelectric bubble domains

Vivasha Govinden[1*], Suyash Rijal[2*], Qi Zhang[1†], Yousra Nahas[2], Laurent Bellaiche[2], Nagarajan Valanoor[1], Sergei Prokhorenko[2†]

[1] *School of Materials Science and Engineering, University of New South Wales, Sydney, NSW 2052, Australia*

[2] *Physics Department and Institute for Nanoscience and Engineering, University of Arkansas, Fayetteville AR 72701, USA*

[*] these authors contributed equally

[†] corresponding authors:
Qi Zhang: peggy.zhang@unsw.edu.au

Sergei Prokhorenko: sprokhor@uark.edu

**Materials and Methods**

*First principles based simulations*

Monte Carlo simulations based on effective Hamiltonian model of PbZr$_{0.40}$Ti$_{0.60}$O$_3$ ultrathin films are used for the theoretical calculations (20). Full details are given in Supplementary Information. Considering epitaxial film geometry, periodic boundary conditions were applied along pseudo-cubic [100] and [010] directions. All simulations were performed for a 56 x 56 x h supercell (h is the thickness of the ferroelectric in unit cells). The following simulation sequence was employed: (i) the system was first thermally quenched from 2000 K to 300 K at zero bias field followed by (ii) the $E_b$ field applied along out-of-plane [00$\bar{1}$] pseudo cubic direction (-z direction), raised from 0 to 0.18 V/nm with 0.02 V/nm steps. 10,000 MC sweeps were performed at each temperature and electric field simulation.

*Film Synthesis*

Two heterostructures, namely PbZr$_{0.4}$Ti$_{0.6}$O$_3$ (PZT-single layer) and PbZr$_{0.4}$Ti$_{0.6}$O$_3$/SrTiO$_3$/PbZr$_{0.4}$Ti$_{0.6}$O$_3$ (PZT-sandwich) films were synthesized on 40 nm thick La$_{0.67}$Sr$_{0.33}$MnO$_3$ (LSMO) buffered (001) oriented SrTiO$_3$ (STO) substrates (Shinkosha, Japan) via pulsed laser deposition (PLD, Neocera, USA). The thickness of PZT single layer is approximately 10 nm, while for PZT/STO/PZT-sandwich, the thickness of top and bottom PZT layers, and STO spacer layer are 9.5±0.5 nm, 9.5±0.5 nm and ~0.8 nm (i.e., 2 unit cells), respectively. The deposition details can be found elsewhere [30].

*Scanning probe microscopy investigation*

The topography and domain patterns of the as-grown state for both PZT-single layer film and PZT-sandwich films were obtained on a commercial scanning probe microscope (Cypher S, Asylum Research, US) using a Pt/Cr coated probe (Multi75GE, BudgetSensors, Bulgaria) under a force of <100 nN. The force constant and free resonance frequency of the probe cantilever are 3 N/m and 75 kHz, respectively.

*Tomographic Atomic Force Microscopy* (TAFM)

Non-destructive mechanical scanning was conducted at film surface under piezoresponse force microscopy (PFM) mode with a low cantilever deflection (with force ≤ 1 μN) using a sharp conductive single crystal diamond probe (AD-40-AS, Bruker, US). The radius and the height of the diamond probe are 10 ± 5 nm and 12.5 ± 2.5 μm, respectively. Force constant and free resonance frequency of the probe cantilever are 40 N/m and 180 kHz, respectively. The topography and PFM (amplitude and phase) images were acquired simultaneously at an AC amplitude of 300 mV. AC bias was set to zero and a force ranging from 7 μN to 15 μN was applied to the film surface using the stiff probe for TAFM scanning. Note that the force was tuned by the cantilever deflection, and under a critical force, the process becomes destructive and triggers the thin film milling. After each TAFM scan, a non-destructive PFM scan was conducted (with force ≤1 μN) at an AC amplitude of 300 mV.

**Numerical simulations and theory details**

**Effective Hamiltonian simulations.** Monte Carlo simulations based on effective Hamiltonian model of PbZr$_{0.40}$Ti$_{0.60}$O$_3$ ultrathin films are used for the theoretical calculations [1]. The corresponding internal energy has the form:

$$\varepsilon_{tot}(\{\mathbf{u}_i\},\{\alpha_i\},\{\mathbf{v}_i\},\eta) = \varepsilon_{Heff}(\{\mathbf{u}_i\},\{\alpha_i\},\{\mathbf{v}_i\},\eta) + \beta\sum_i\langle\mathbf{E}_{dep}\rangle\cdot Z^*\mathbf{u}_i - \sum_i \mathbf{E}_b\cdot Z^*\mathbf{u}_i, \quad (1)$$

where $\mathbf{u}_i$ is the local soft mode in the i$^{th}$ unit cell (u.c.) of the film. The product of $\mathbf{u}_i$ with the local mode Born charge $Z^*$ and the lattice parameter $a_0$ gives the local electric dipole in the i$^{th}$ cell. The atomic distribution of the B-site sublattice in the alloy, where the B-site is either Ti or Zr, is characterized by integer lattice variables $\alpha_i$. Particularly, $\alpha_i = 1(\alpha_i = -1)$ indicates B-site is occupied by Ti (Zr) atom. Variable $\{\mathbf{v}_i\}$ describes inhomogeneous strain inside the film. $\eta$ is the homogeneous strain tensor. As the films are epitaxially strained, three of the six components of $\eta$, when expressed in Voigt notation, are always kept fixed, namely $\eta_1 = \eta_2 = \delta$ and $\eta_6 = 0$ where $\delta$ is the strain due to the lattice mismatch. The rest of the components are allowed to relax. The first term, $\varepsilon_{Heff}$, represents the alloy effective Hamiltonian which is the intrinsic energy of ferroelectric film [2]. Here, the treatment of dipole-dipole interaction [3] assumes ideal open circuit (OC) condition, *i.e.,* maximum depolarizing field ($\langle\mathbf{E}_{dep}\rangle$). The second term $\beta\sum_i\langle\mathbf{E}_{dep}\rangle\cdot Z^*\mathbf{u}_i$ represents screening of depolarizing field where the extent of screening is given by the factor $\beta$ (screening parameter). The combination of the first and second term leads to the final form of depolarizing field given by $(1-\beta)\mathbf{E}_{dep}$. Here, $\beta = 1$ represents ideal short-circuit condition (*i.e.,* no depolarizing field), and $\beta = 0$ represents ideal open-circuit condition (*i.e.,* maximum depolarizing field). Open-circuit (short-circuit) boundary condition, $\beta$=0 ($\beta$=1), is reported to stabilize labyrinthine/stripe/polydomain (monodomain) states. Varying $\beta$ from 0 to 1 results in a transition from labyrinthine/striped domains into monodomain phase [1], passing through a myriad of topological structures [4], one of them being polar bubble phase. The third term imitates the effect of built-in/ intrinsic bias, present in the experimentally grown samples. In our simulations, we varied both $\beta$ and external electric field $\mathbf{E}_b$. An epitaxial strain of -2% was considered for all the simulations. The effect of externally applied stress is described by the bi-linear term $-a_0^3\sum_\gamma \sigma_\gamma \eta_\gamma$ where the sum goes over the components $\gamma$ of the stress ($\sigma_\gamma$) and homogeneous strain tensors ($\eta_\gamma$).

Considering epitaxial film geometry, periodic boundary conditions were applied along pseudo-cubic [100] and [010] directions. All simulations were performed for a $56 \times 56 \times h$ supercell ($h$ is the thickness of the ferroelectric in unit cells). The following sequence was employed: (i) supercell quenched from 2000 K to 300 K at zero E field, followed by (ii) $E_b$ field applied along out-of-plane [00$\bar{1}$] pseudo cubic direction (-z direction), raised from 0 to 0.18 V/nm with 0.02 V/nm steps. 10000 MC sweeps were performed at each temperature and electric field step which is sufficient in terms of structural relaxation.

**Conservation of the depolarizing field magnitude.** Up to the leading order, the mean-field Landau potential $F$ of the PZT layer can be written as

$$F = \kappa\, P_z^2 + E_d^0 P_z - E_b P_z + O(P^4), \qquad (2)$$

where the $\kappa$ term describes the free-energy gain related to ferroelectric instability ($\kappa < 0$). $E_d^0$ stands for the bare depolarizing field magnitude. The second and third terms thus describe the energy penalty associated with the emerging depolarizing field and the energy gain due to built-in/external bias $E_b$, respectively. The $P_z = 0$ state in Eq. (2) corresponds to an inhomogeneous labyrinth domain pattern rather than the usual paraelectric phase. The omission of higher order terms is justified by the domination of the bare depolarizing field energy over the $\kappa$ term which here holds, to a good accuracy, for all modulated phases (e.g., labyrinth, disconnected labyrinth, bimeron and bubble states) [5]. At the same time, it should be noted that approximation described by Eq. (2) breaks in the direct vicinity of the transitions (labyrinth-bubbles and bubble-monodomain) as well as for the monodomain state as, in both cases, non-linear effects become important.

From Eq. (2) one can readily obtain the equation of state for the out-of-plane polarization

$$2\,\kappa\, P_z + E_d^0 - E_b = 0, \qquad (3)$$

where $E_d^0$ can be approximated [3] as

$$E_d^0 \approx (1-\beta)\frac{\Delta(h)}{\varepsilon_\infty}P_z. \qquad (4)$$

Here, $\varepsilon_\infty$ denotes the optical dielectric permittivity and $\Delta(h)$ takes into account the weak dependence of the depolarizing factor on the film thickness $h$. Then, substituting Eq. (4) in Eq. (3) yields

$$\left(2\kappa + (1-\beta)\frac{\Delta(h)}{\varepsilon_\infty}\right)P_z = E_b. \qquad (5)$$

Noting that $\left(2\kappa + (1-\beta)\frac{\Delta(h)}{\varepsilon_\infty}\right)$ is the inverse susceptibility $\varepsilon^{-1}$ of the film, we finally obtain $P_z/\varepsilon = E_b$, or, equivalently

$$E_d = E_b, \qquad (6)$$

where $E_d = P_z/\varepsilon$ denotes the residual depolarizing field magnitude. The residual field $E_d$ is smaller than the bare depolarizing field $E_d^0$ since the former takes into account the screening of polarization-induced bound charges by the polarization self-response to $E_d^0$. For instance, the bare field vanishes at $\beta = 1$, while the residual field $E_d$ becomes equal to zero at $\beta_{SC} = 1 + \frac{\varepsilon_\infty}{\Delta(h)}2\kappa < 1$.

Finally, we analytically derive the dependence of $P_z$ on the screening strength. This case is particularly simple since $\beta$ enters explicitly in Eq. (5). Solving Eq. (5) gives

$$P_z = \frac{\varepsilon_\infty E_b}{\Delta(h)}(\beta_{SC} - \beta)^{-1}. \tag{7}$$

This formula is used to fit $P_z(\beta)$ dependence in **Fig. 1g**. In order to test the conservation of residual depolarizing field we also plot in **Fig. S1** the dependences of $E_d = P_z/\varepsilon$ on the stress, the screening $\beta$ and the thickness. For this, the susceptibility $\varepsilon$ is calculated from molecular dynamics (MD) simulations using the fluctuation-dissipation theorem [6]. The same molecular dynamics trajectories are used to compute the out-of-plane polarization $P_z$. In both case, 50000 MD steps of 0.5 fs are used to compute the averages. Such approach allows to reduce the computational effort, but, in turn, gives rise to significant statistical errors in the computed $E_d$ values. Despite such errors, the $E_d = E_b$ trend can be clearly seen for both the changing σ and $\beta$ cases. **Figure S1b** also shows the aforementioned breaking of the $E_d$ conservation at the transition from the bubble domains to a homogeneously polarized state. In the latter case, the calculated $E_d$ value is much lower than $E_b$ as a result of the high susceptibility of the monodomain pattern at β > $β_{SC}$. For this range of screening conditions, all dipoles point in the same direction, but their equilibrium magnitude is still low when compared to the ideal $\beta = 1$ short-circuit case (**Fig. 1g**). Therefore, the applied bias can easily increase the average dipole magnitude which explains the high ε and, consequently, low $E_d$ values. In case of varying thickness, the mismatch between the calculated $E_d$ and $E_b$ increases with decreasing thickness. Such trend can be explained by the growth of the error in $\varepsilon$ with decreasing thickness due to the finite-size scaling [6].

**Experiments**

**Tomographic Atomic Force Microscopy.** The typical TAFM-PFM process is shown in **Fig. S2a**, taking PZT-sandwich sample as an example. After an initial PFM scan, TAFM is conducted over the white box (600 nm × 300 nm) under 7 μN mechanical force for 5 to 6 times. Then PFM scanning is performed to image the domain for the as-obtained tomographic surface. **Fig. S2b** describes how a depression is created after TAFM and a tomographic surface emerges. **Fig. S2c** plots the cross-sectional profile of the tomographic surface (along the white dashed arrow from point A to B in **Fig. S2d**) to show their depth below surface of as-grown film. The top half of the mechanical scanned rectangle region in **Fig. S2d** has experienced one more 7 µN TAFM scan than the bottom half; this allows us to study the role of a gradual change in depth. It is found that each 7 µN scan causes approximately 4 ± 2 nm depletion. After the 6th scan, the depth of the tomographic layer reaches 24 ± 2 nm which is already below the bottom PZT layer. The scan rate was fixed at 1 Hz for all scans. Similar TAFM-PFM procedure is applied for PZT-single layer.

**Reference sample.** **Fig. S3** shows the as-grown domain configuration of reference sample PZT single layer.

**Domain configuration of PZT/STO/PZT-sandwich films. Fig. S4** shows the topography and domain patterns of as-grown PZT/STO/PZT-sandwich films which was captured under <100 nN force. This mechanical force has negligible influence on the domain pattern during scanning [7]. Labyrinthine domains patterns are observed for the as-grown PZT thin films with a domain width of approximately 30±5 nm. The labyrinth and matrix phases show a clear 180° phase difference indicating their domain polarization to be in opposite directions. Note here the black/purple and yellow contrast phases indicate upward and downward polarization, respectively.

With the scanning force increased to ~1 µN, the upward labyrinthine domain width slightly decreases to 20±5 nm (**Fig. S4b**), suggesting downward domain switching under the mechanical force [8]. Note that no topography change is observed, suggesting that the transitions are not correlated with either surface damage or ferroelastic switching. Next, a scan was conducted in the centre (600 nm × 300 nm white rectangle box in **Fig. S4c**) of previously 1 µN scanned area with an enhanced mechanical force (gradually increased from 4 µN to 6 µN from bottom to top). A further domain transition from narrow labyrinthine pattern towards nanoscale bubble-like domains is observed within this 600 nm × 300 nm region. All the as-grown poled-up labyrinthine domains are broken down into nanoscale patterns with an average planar domain size of approximately 10 nm as shown in their amplitude images. No clear phase contrast can be observed for these evolved nanoscale domains and the whole mechanically scanned area shows the same phase contrast as the poled down matrix (yellow contrast). A similar observation has been previously reported for bubble domains where the fuzzy domain contrast is attributed to non-Ising domain walls [7, 9]. Note the sharp contrast in the domain amplitude patterns outside white box excludes the possibility of any tip condition change.

This reveals a simple route of acquiring stable nanoscale bubble domains by applying external mechanical force on as-grown labyrinthine domains. For the previously reported nanoscale bubble domains [7], the ultrathin tetragonal PbZr$_{0.2}$Ti$_{0.8}$O$_3$ films (7 nm thick) are subjected to a compressive strain. The desired mechanical and electrical boundary conditions of the films are achieved by tuning depolarization field using an additional STO spacer layer [7]. In this study, a similar PZT/STO/PZT sandwiched structure is used [10], but with much thicker PZT layers (total thickness of ~20 nm) which are almost fully relaxed (**Fig. S5**). Thus, only conventional c-domains are observed in the as-grown state. Under external mechanical force applied using the scanning probe microscopy (SPM) probe, the strain condition of the film starts to change [8]. When this strain meets the mechanical boundary condition criteria for topological domains, the conventional domains evolve into the nanoscale domains. This domain transition from labyrinthine to nanoscale domains under mechanical scanning brings about an exciting possibility that the topological transition might be triggered by a mechanical force.

**Crystallographic structures of PZT-single layer and PZT/STO/PZT-sandwich heterostructures. Fig. S5a-b** show the XRD patterns of PZT-single layer and PZT/STO/PZT-sandwich heterostructures, respectively. PZT layer from both films are in a near-relaxed states, with the out-of-plane lattice parameter of 0.412 nm (PZT-single layer) and 0.414 nm (PZT/STO/PZT-sandwich), respectively.

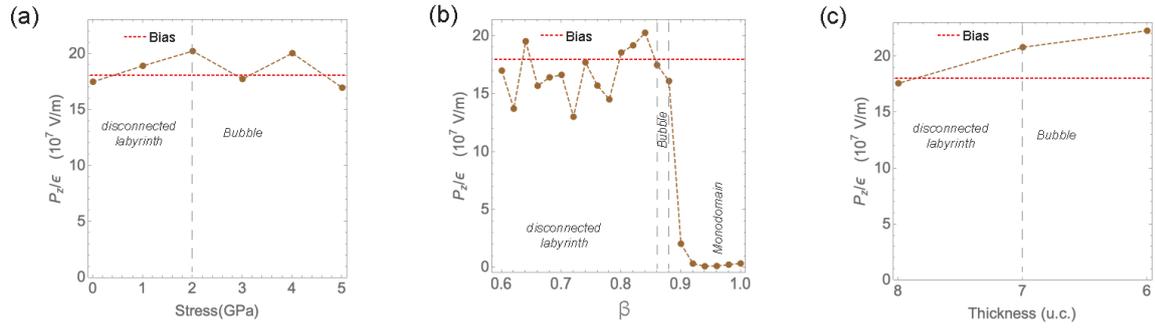

**Fig. S1.** (a) Computed dependence of the residual depolarizing field magnitude $E_d = P_z/\varepsilon$ on the (a) stress for β=0.86 and h=8 u.c. (b) screening β for σ =0 GPa and h=8 u.c. and (c) thickness at β=0.86 and σ =0 GPa. The dashed gray lines indicate the transitions between characteristic domain patterns while the dashed red lines indicate the applied bias level.

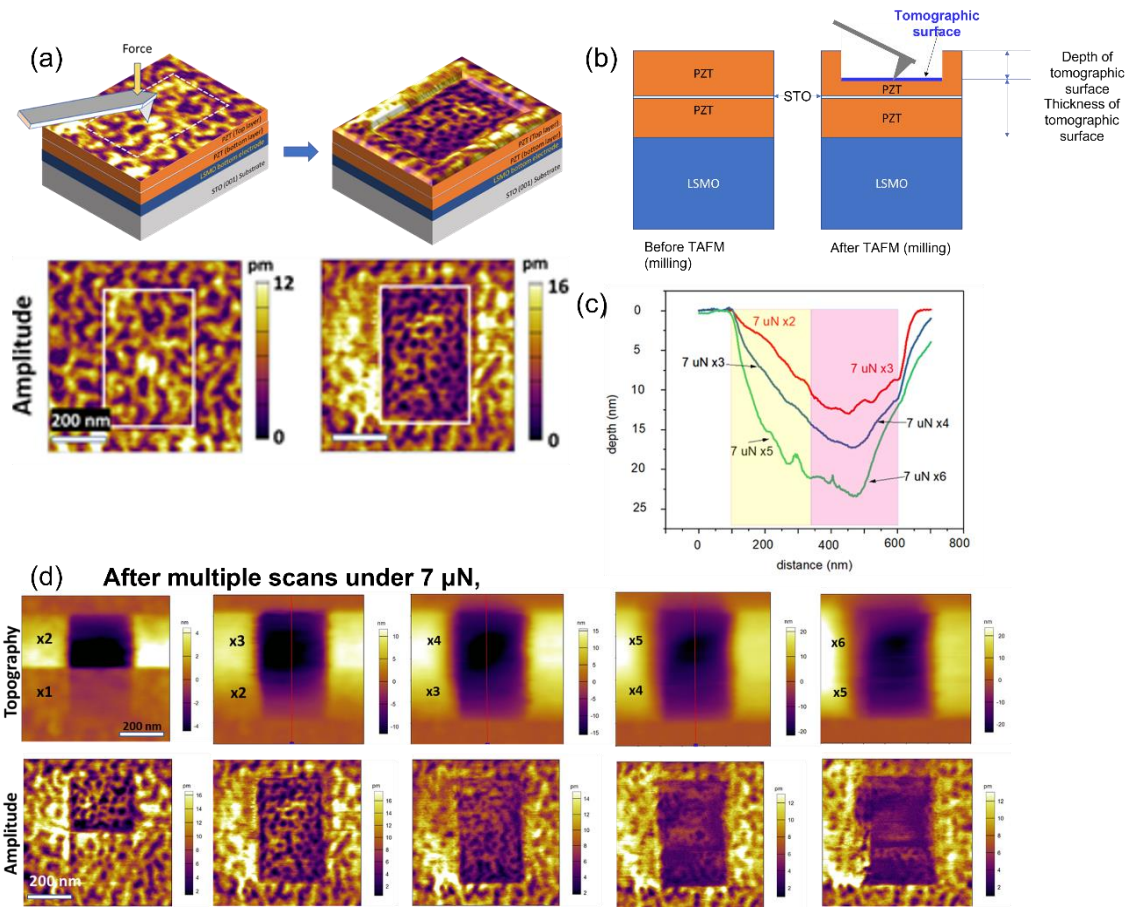

**Fig. S2.** (a) Schematic of TAFM-PFM procedure for PZT-sandwich film; TAFM scanning is conducted over the white box, followed by a PFM scanning to collect the domain information (e.g., PFM amplitude) at the as-obtained tomographic surface. (b) Definition of tomographic surface, and depth and thickness of tomographic surface. (c) Cross sectional profile of the force scanned area, showing the depth of concaved region changing with the scanning times under 7 µN. (d) topography change with the TAFM scanning under 7 µN for multiple times. The numbers in (d) indicate the number of scanning times under 7 µN in the rectangular zone. the top half area is subject to an extra scan, as compared to bottom half.

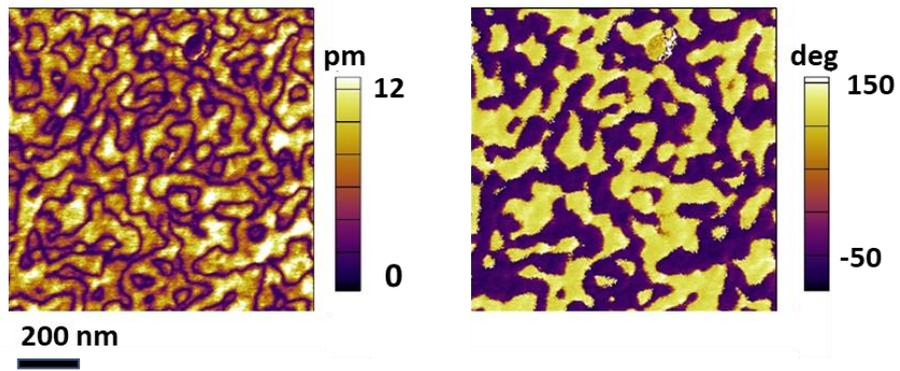

**Fig. S3.** Amplitude (left) and phase (right) images of PZT-single layer measured under 100 nN.

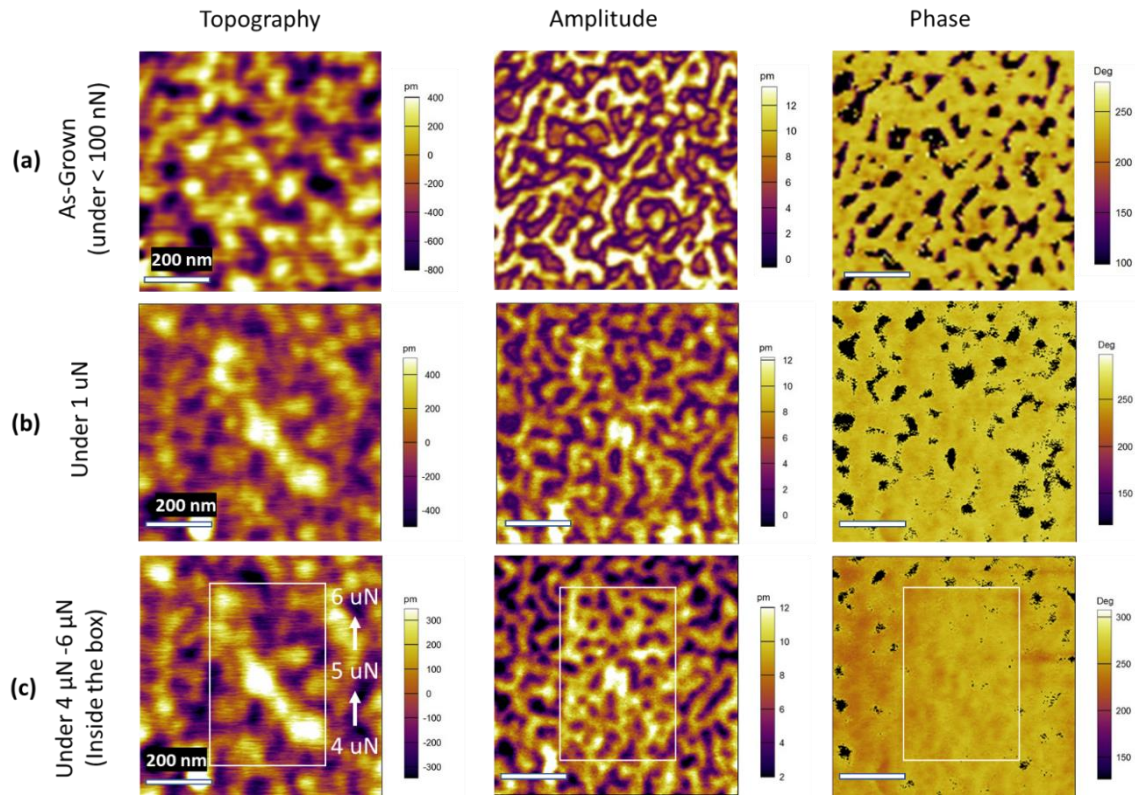

**Fig. S4.** Topography and PFM amplitude and phase images of PZT/STO/PZT-sandwich films under (a) <100 nN force, (b) 1 µN force and (c) 4-6 µN force. Note 4-6 µN force was applied to the rectangular region (in white box) during scanning followed by PFM scanning under 1 µN force. The force was increased from 4 µN to 6 µN gradually from bottom to top. No topography change was observed after force application indicating no milling has been occurred under this force. Nevertheless, a domain evolution from labyrinthine domains to a nanoscale bubble-like domains was observed.

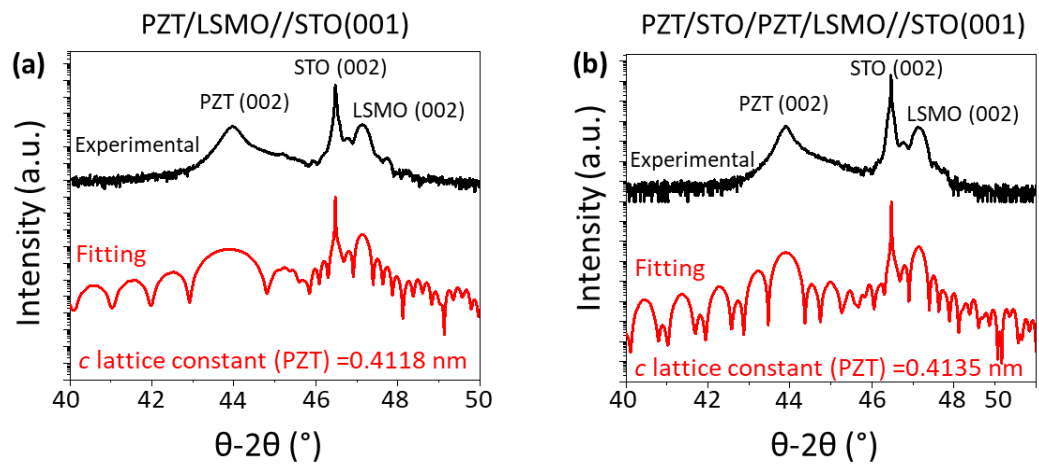

**Fig. S5.** Simulated and Experimental X-ray diffraction data of (a) PZT-single layer and (b) PZT/STO/PZT-sandwich films.